# On Capturing the Spreading Dynamics over Trading Prices in the Market

Hokky Situngkir
[hs@compsoc.bandungfe.net]
Dept. Computational Sociology
Bandung Fe Institute

**Abstract**
While market is a social field where information flows over the interacting agents, there have been not so many methods to observe the spreading information in the prices comprising the market. By incorporating the entropy transfer in information theory in its relation to the Granger causality, the paper proposes a tree of weighted directed graph of market to detect the changes of price might affect other price changes. We compare the proposed analysis with the similar tree representation built from the correlation coefficients of stock prices in order to have insight of possibility in seeing the collective behavior of the market in general.

## 1. Introduction

Market is the place where information flows of interacting agents, investors, firms. Information flows spread in many channels, be it micro-interaction among market actors or the changes of prices. Decisions made only from one side of observation betray the complex nature of the market and is doomed to fail. Works on market dynamics observe levels by levels of the market [13], from the micro level of investors and how they interact shaping the market price, and also the works on the statistics of market in order to gain comprehensive view on the dynamics of a single price and the spectrum of stocks traded in the market. Observing the spectrum of the stock prices are actually looking over the movement of prices as they perceived by investors and market actors from their portfolio's dashboard, strategies, and various kinds of decision making tools.

Stock price movements are interdependent one another, for investors of a particular stock would never close their eyes with the other stock prices, or at least the composite index tried to reflect the global situation of the market. This is obvious from the stylized statistical facts of the market. The "almost" symmetry and fat tailed distribution of returns are somehow demonstrated the interdependent nature of the market. The return of the stock prices, $r(t) \in \Re$,

$$r(t) = \ln \wp(t) - \ln \wp(t-1) \qquad (1)$$

where $\wp(t)$ denotes the price at time $t$, are clustered in its volatility for the way investors and traders behave towards the market changes [2].

One interesting discussions about the stock market dynamics are the way information flows in the market as reflected in changes of one price with the other. It is not a secret that an investor frequently decide her actions toward a particular stocks by referring to other stock's price changes. Thus, there is a sort of information flows in timely manner from stocks to stocks price changes. The changes of one stock prices may affect the changes of the other ones. Statistically, the question is how we can measure it, even though some responds might come from our understanding over the information theory.

Entropy is a notion in information theory about disorderliness of a system based upon the micro-states comprised it. The changes of entropy are related to the timely dynamics of the system. When information flows, the entropy is transferred and the system is changes over time. A change over time in a system does not have to be related to the cause and effect within the system, though. A thing occurs and is followed by other changes does not have to be seen the cause of the other changes. Nonetheless, the spreading of information can be seen as a thing causing other thing, especially when it comes to highly interacting micro states within a system.

The study of entropy transfer is related to information flows over the changing of micro-states within the interacting composite of the system. The paper is the discussion about this in our observation of the Indonesian stock market. The structure of the paper is as follows. The information theoretical concept of entropy transfer is introduced as we see stock price moves from one another. The concept is also related to the causality concept as introduced by Granger-causality [5]. The paper tries to portray the spreading dynamics over the stock prices within a market by the representation of weighted directed tree and compare the result with the one yielded with correlation coefficients among stock prices [7]. The discussion is about how we can gain more information about the market on the representation of the tree, be it from the information theoretic entropy transfer and Granger-causality and also from the ultrametric tree yielded from the analysis of ultrametric tree.

## 2. Model

Market time series can be seen as a composite of the set of $\mathcal{M}$ interacting dynamical sub-system. Investors put their trading decisions due to their portfolio and market strategies, shaping the prices of the traded stocks. Over time, the prices are depicted the dynamical processes within the collective behavior of the investors. The vicissitudes of a price could affect the dynamic of other prices due to their portfolios. Capturing the dynamics of spreading ups and downs within the market is observing the information flow from one price to one another. For instance we have a source system $\mathcal{Y}(t)$ as the source of information affecting other sub-system $\mathcal{X}(t)$, collecting the remaining sub-systems in the vector of $\mathcal{Z}(t)$. From the information theoretic studies [4, 14], we know that the differential entropy of a random vector $\mathcal{X}$ is defined,

$$h(\mathcal{X}(t)) = -\int_{\mathfrak{R}^d} p(\boldsymbol{x}) \ln p(\boldsymbol{x})\, d\boldsymbol{x} \quad (2)$$

as the random vector takes value in $\mathfrak{R}^d$ with probability density function $p(\boldsymbol{x})$. When the random variable $\mathcal{X}(t)$ is multivariate discrete of all possible values of $x \in \{x_1, x_2, \dots, x_n\}$, the entropy is

$$H(\mathcal{X}(t)) = -\sum_{i=1}^{n} p(x) \ln p(x_i) \quad (3)$$

where now, $p$ is the probability mass function of $\mathcal{X}$.

Thus, the transfer entropy, $\mathcal{T}_{Y(t)\to X(t)|Z(t)}$, of the previous $\mathcal{X}(t)$, $\mathcal{Y}(t)$, and $\mathcal{Z}(t)$ is written as,

$$\mathcal{T}_{Y(t)\to X(t)|Z(t)} = H(X(t)|[\![X^-(t), Z^-(t)]\!]) - H(X(t)|[\![X^-(t), Y^-(t), Z^-(t)]\!]) \quad (4)$$

where $H(A)$ denotes the entropy of the variable $A$, $H(A|B)$ the conditional entropy,

$$H\big(X(t)|Y(t)\big) = -\sum_{i=1}^{n}\sum_{j=1}^{m} p(x_i, y_i) \ln p(x_i|x_i) \quad (5$$

for $m$ can be different with $n$, and $p(x_i|x_i)$ as the conditional probability, as to

$$H\big(X(t), Y(t)\big) = -\sum_{i=1}^{n}\sum_{j=1}^{m} p(x_i, y_i) \ln p(x_i, x_i) \quad (6)$$

with $p(x_i, x_i)$ as the joint probability. The past of vectors $\mathcal{X}(t)$, $\mathcal{Y}(t)$, and $\mathcal{Z}(t)$ are respectively $X^-(t) = \{X(t-1), X(t-2), \dots, X(t-p)\}$, $Y^-(t) = \{Y(t-1), Y(t-2), \dots, Y(t-p)\}$, and $Z^-(t) = \{Z(t-1), Z(t-2), \dots, Z(t-p)\}$ with the length vector $p$, and the vectors in the bracket $[\![A, B]\!]$ are concatenated.

From there we have,

$$\mathcal{T}_{Y(t)\to X(t)|Z(t)} \equiv \sum p\big(X(t), X^-(t), Y^-(t), Z^-(t)\big) ln \frac{p\big(X(t)|X^-(t), Y^-(t), Z^-(t)\big)}{p\big(X(t)|X(t), Z^-(t)\big)} \quad (7)$$

where $p(A)$ is the probability associated with the vector variable $A$, and $p(A|B) = \frac{p(A,B)}{p(B)}$, the probability of observing $A$ with knowledge about the values of $B$.

The notion of the entropy is an information theoretic terminology that can be regarded as the measure of the disorder level within the random variable of the time series data. Transfer entropy from $\mathcal{Y}(t)$ to $\mathcal{X}(t)$ is reflecting the amount of disorderliness reduced in future values of $\mathcal{X}(t)$ by knowing the past values of $\mathcal{X}(t)$ and the given past values of $\mathcal{Y}(t)$. Time "moves" as entropy is transferred and observed in flowing information from series to series.

We have two regressions toward $X(t)$, the first is the moving series without putting the $Y(t)$ into account,

$$X(t) = A[\![X^-(t), Z^-(t)]\!] + \in_1 (t) \tag{8}$$

and the other one which regard to the information transfer from $Y(t)$ to $X(t)$,

$$X(t) = A[\![X^-(t), Y^-(t), Z^-(t)]\!] + \in_2 (t) \tag{9}$$

where A is the vector of linear regression coefficient, and the $\in_1$and $\in_2$ are the residuals of the regression. The residuals have respective variances of $\sigma(\in_1)$ and $\sigma(\in_2)$, and under Gaussian assumption, the entropy of $X(t)$ is,

$$H(X(t)|\, X^-(t), Z^-(t)) = \frac{1}{2}(\ln \sigma(\in_1) + 2\pi e)) \tag{10}$$

and

$$H(X(t)|\, X^-(t), Z^-(t)) = \frac{1}{2}(\ln \sigma(\in_2) + 2\pi e)) \tag{11}$$

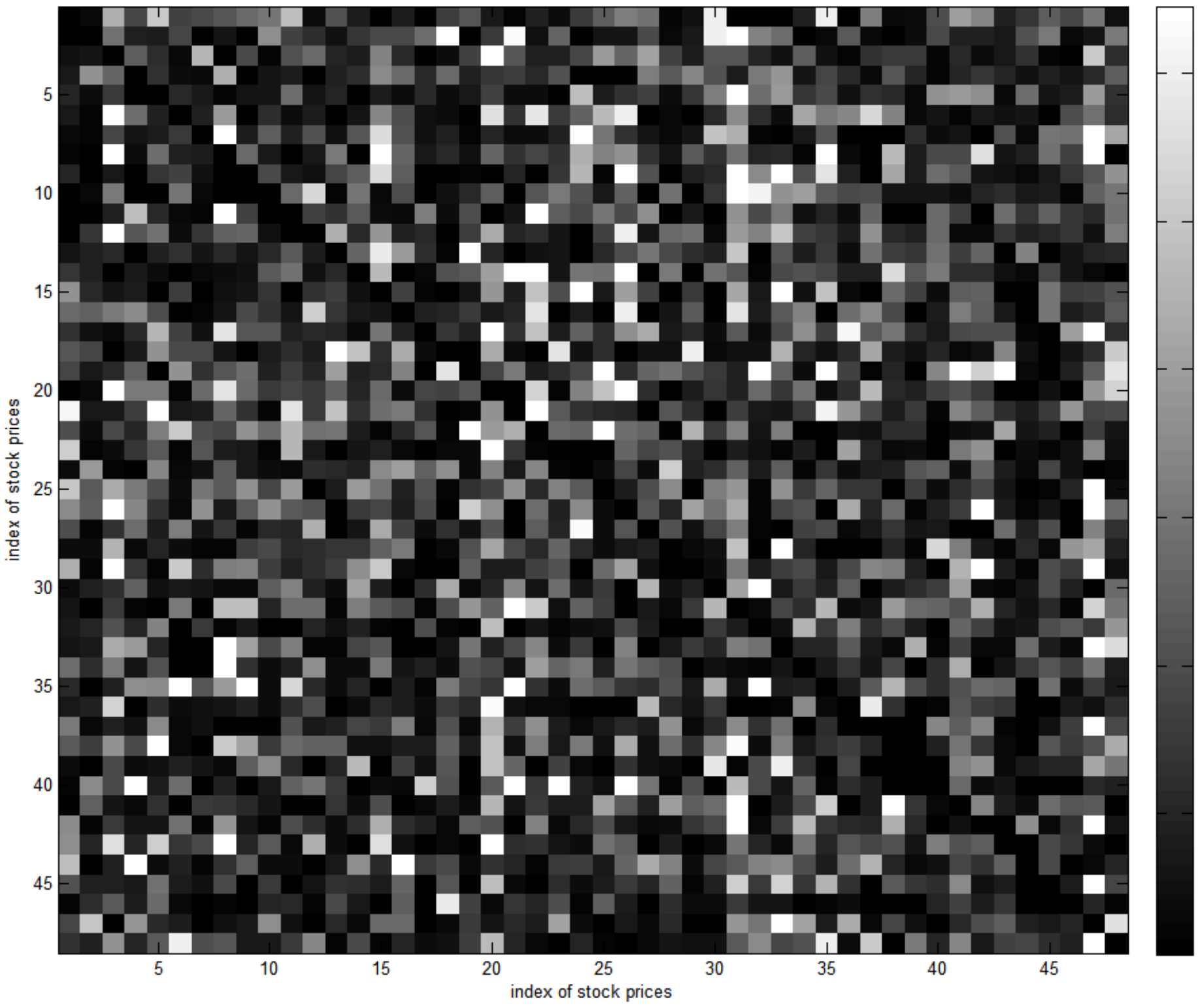

**Figure 1.** The representation of calculated cause and effect or information transfer among most traded stocks in Indonesia Stock Market for 3 years of trading (2012-2015).

Thus, we can get the estimated transfer entropy

$$\mathcal{T}_{Y(t)\to X(t)|Z(t)} = \frac{1}{2}\ln\frac{\sigma(\epsilon_1)}{\sigma(\epsilon_2)} \tag{12}$$

This information theoretic notion opens the bridging discussions to the statistics of the autoregressive methods of Granger-causality [5, 6]. The idea of Granger-causality came from understanding that $\mathcal{Y}(t)$ is said to cause $\mathcal{X}(t)$ for $\mathcal{Y}(t)$ helps predict the future of $\mathcal{X}(t)$ [5, 8, 9]. This is a statistical concept equivalent (with some notions) with the transfer entropy [1, 3], of which in our case, the Granger-causality is estimated as,

$$\mathcal{G}_{Y(t)\to X(t)|Z(t)} = \ln\frac{\sigma(\epsilon_1)}{\sigma(\epsilon_2)} = 2\,\mathcal{T}_{Y(t)\to X(t)|Z(t)} \tag{13}$$

Thus, the entropy transferred can be seen as causal relations among random variables, with which we can learn the spreading dynamics over trading prices in the market represented by the multivariate data.

### 3. Experiment on Data

After reshaping the return of the time series data, *i.e.*: by detrending and remove the temporal mean from all time series [2, 15], the set of the data is analyzed for the information theoretic causal estimations. Market is a complex system with interacting agents through the market prices, thus in so many ways, causality may be detected even at a small glimpse of ups and downs of a market object. This is shown in a scale of the $n \times n$ matrix in figure 1.

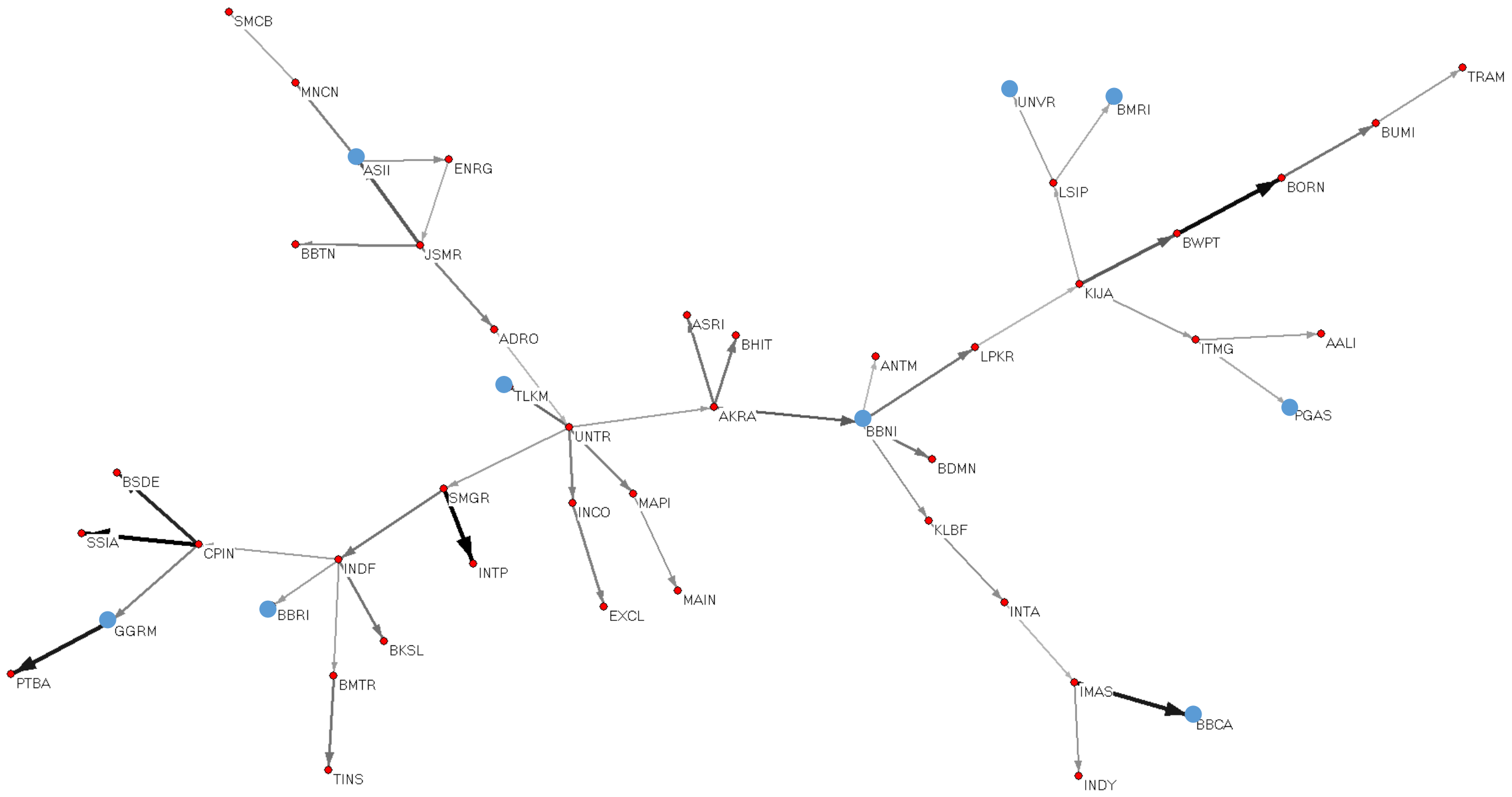

**Figure 2.** Calculated spreading dynamics over the years of trading 2012-2015 among some popular Indonesian stocks, the thicker the arc, the bigger the detected information flows within respective price time series.

Therefore, in order to observe the dynamics of spreading information as relatively causal price changes, we depicted the calculated yields as directed weighted graph. The directed weighted graph, defined as $G = (V, A)$, as they are made by the set of vertices, $V$, and the set of arcs, $A$. The graph is represented as adjacency matrix with elements $a_{ij} \in A$, where the $i, j \in V$. In the representation, we show the weight of the arcs as the maximum causality index,

$$a_{ij} = \begin{cases} \mathcal{T}_{i\to j}, p(\mathcal{T}_{i\to j}) \geq \sum_{j=1}^{p} \mathcal{T}_{i\to j} \\ 0, otherwise \end{cases} \tag{14}$$

From here, we have in hand the mapping of the spreading price changes in the stock markets, in terms of causality. For 3 years of trading period (2012-2015), we depicted the result as shown in figure 2. The figure represents the collective dynamics of the ups and downs of stock prices traded in the Indonesian Stock Market within 3 years of trading.

In the figure it is also shown some stocks with highest market capitalization, marked with blue nodes. It is interesting to see that the highest capitalization stocks are spreading over the leaves within the tree. When it comes to the dynamics of information (entropy) transfer, somehow it can reflect the structures of most portfolios and strategies used by the investors due to the period. The figure demonstrates a bird's eye view on the collective dynamics of price movements in the market. Nonetheless, many studies related to the stocks portfolio due to the fundamental aspects [12] of the observed firm can gain from the mapping in the figure for predictive assignments.

**4. Discussions: on relations with ultrametric tree of stock prices**

The observation on the spectral price movements in the stock market, by employing the "mapping" using the similar hierarchical tree has also been proposed previously [7, 11]. The idea is to transform the correlation coefficient into a special space metric, name ultrametric space by,

$$d_{ij} = \sqrt{2(1 - c_{ij}} \tag{15}$$

where $d_{ij}$ is the distance from the respective correlation coefficient $c_{ij}$ between stock $i$ and $j$. The unique space metric complies the properties of the Euclidean space,

$$properties \to \begin{cases} d_{ij} = 0, i = j \\ d_{ij} = d_{ji} \\ d_{ij} \leq d_{ik} + d_{kj} \end{cases} \tag{16}$$

with a special properties of

$$d_{ij}^{ultrametric} \leq \max\left(d_{ik}^{ultrametric}, d_{kj}^{ultrametric}\right) \tag{17}$$

By using the Kruskal algorithm to have the Minimum Spanning Tree [7], the tree representation of correlative behavior among stock prices is delivered. In order to compare this representation with the spreading dynamics shown in figure 2, the same data is processed by employing the equation 15. The yielded tree is shown in figure 3.

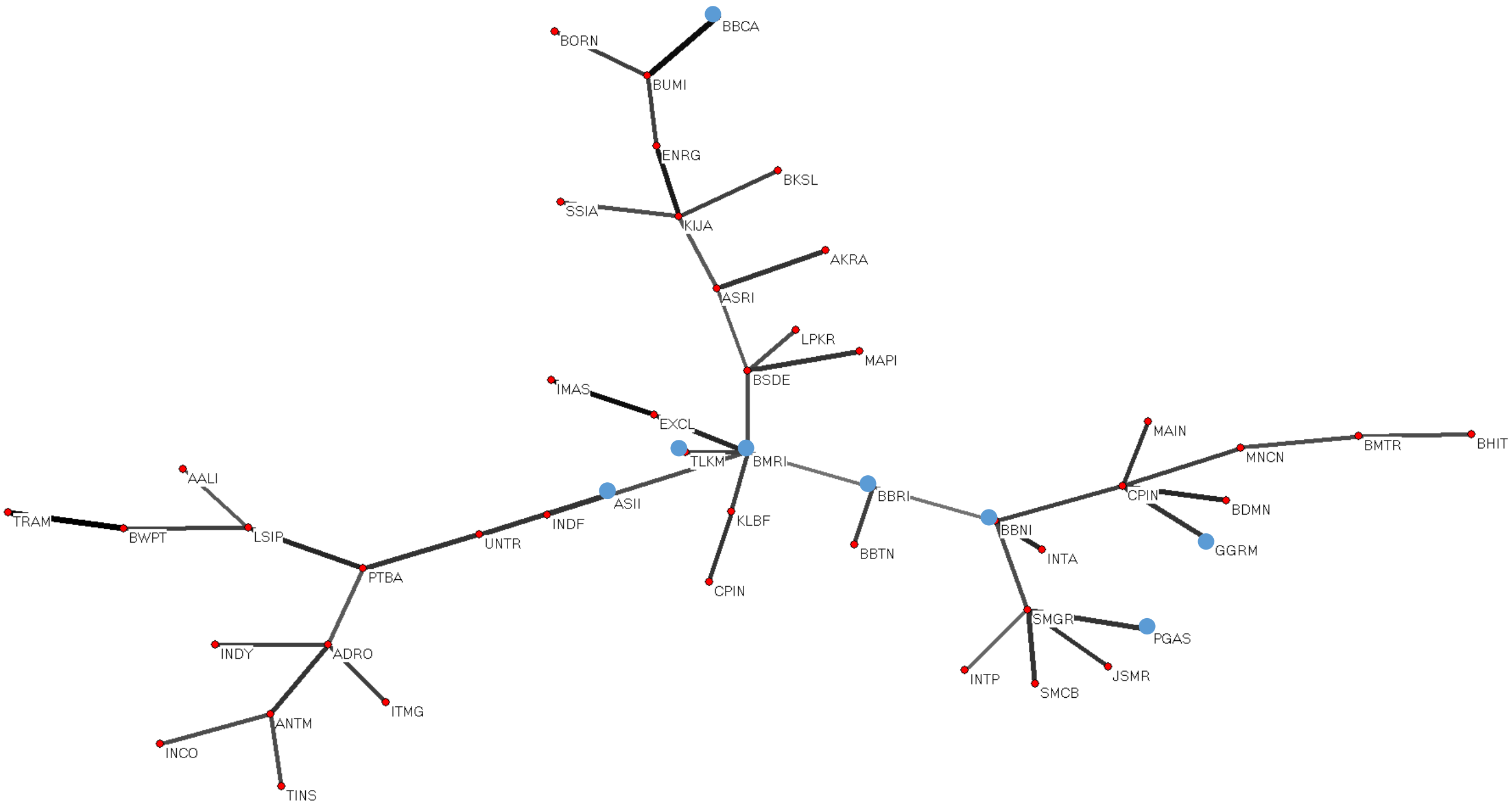

**Figure 3.** The Minimum Spanning Tree for some stocks in Indonesian Stock Market (2012-2015).

The tree of spreading information dynamics and ultrametric tree of stock prices are both representing in tree visualization but capturing different aspects of the spectrum of prices in the market. The first captures the information transfer from one stock price to another, by focusing more on the two stock prices with conditionals with the other. The ultrametric tree is constructed more on whole spectrum of the prices. That is why both represent different information on the observation. For instance, the stocks with highest capitalization stocks are more to become the anchor in the minimum spanning tree of price correlation. They are represented to be the "market mover" within the Indonesian Stock Market for their liquidity. Thus the observation of the correlation tree are supposed to be more on the emerged clustering of the stocks.

It is different with the tree of the spreading information as we proposed in the paper. The tree shows more on some statistical scenario of the dynamics of the market over time. The directed arcs from nodes to nodes shows how information are moves as reflected by the strategies and portfolios of the investors statistically. The "anchor" stocks are shown more likely to be in the outer leaves of the whole tree for their liquidity in the market. Their price dynamics are shown to be more caused rather become cause of other price dynamics. Therefore, the tree of the spreading price dynamics represents a more detailed micro states of the market than the minimum spanning tree.

However, in the sense of investment strategies and portray of the whole market dynamics, both are enriching one another. The ultrametric tree can be seen as a way to construct portfolio and the clustered movements of the price, while the tree of the spreading information gives the aggregated or macro-view of the portfolios used by investors globally.

## 4. Concluding Remarks

The concept of causality as derived from the transfer entropy is introduced to see the dynamics of spreading information among the changes of the stock prices in Indonesia. It is interesting to have a tool of observation to see over changes of stock prices that can be related by detection of the entropy transfer or "causal" aspects. From the resulting "causal-tree" we can see the spreading information dynamics over the changes of price movements. The global view gives insights on how investors collectively perceive the changes of a price and behave towards it. Information flows in the market are detected by the mapping of the detected dynamical relations between one stock with another via their prices.

The discussions also brings on the works on the representation of correlations among stock prices in the ultrametric space of minimum spanning tree. While correlation dynamics as portrayed in the ultrametric space is more about the clustering aspects of price movements, it is demonstrated that the spreading dynamics is observed better by using the proposed observation with the information theory. However, the interplay between both representation may gain more information about the behavior of the market in general.

**Acknowledgement**

The author thank Rolan Mauludy Dahlan for discussion at the early stage of the research presented in the paper. All faults remain author's.